\documentclass[11pt]{article}
\usepackage{amsfonts}
\def \sech{\mathop{\rm sech}\nolimits}

\usepackage{amssymb,amsmath,mathrsfs}
\usepackage{graphicx,picins}
\usepackage{subfigure}
\usepackage{float}
\usepackage[numbers,sort&compress]{natbib}

\allowdisplaybreaks[4]

\begin{document}
\title{\bf{From discrete nonlocal nonlinear Schr\"{o}dinger equation to coupled discrete Heisenberg ferromagnet equation} }
\author{ Li-Yuan Ma$^{a}$, Shou-Feng Shen$^a$ and Zuo-Nong Zhu$^b$\footnote{Corresponding author. Email: znzhu@sjtu.edu.cn; mathssf@zjut.edu.cn; mly2016@zjut.edu.cn}
\\
$^a$ Department of Applied Mathematics, Zhejiang University of Technology,\\ Hangzhou 310023, P. R. China\\
$^b$ School of Mathematical Sciences, Shanghai Jiao Tong University,\\ 800 Dongchuan Road, Shanghai, 200240, P. R. China}
\date{ }
\maketitle

\begin{abstract}
 In this paper, we show that the nonlocal discrete focusing nonlinear Schr\"{o}dinger (NLS) and nonlocal discrete defocusing NLS equation are gauge equivalent to the discrete coupled Heisenberg ferromagnet (HF) equation and the discrete modified coupled HF equation, respectively. Under the continuous limit, the discrete coupled HF equation and the modified discrete coupled HF equation lead to the corresponding coupled HF equation and modified coupled HF equation. This means that the nonlocal focusing and defocusing NLS equations are gauge equivalent to the coupled HF and coupled modified HF equations. The solution of the modified coupled HF equation is obtained by using the solution of nonlocal defocusing NLS equation. 
\end{abstract}

\section{Introduction}
\qquad Very recently, Ablowitz and Musslimani proposed an integrable nonlocal NLS equation \cite{4,5,6}
\begin{equation}\label{1}
i q_t+q_{xx}\pm 2qq^{*}(-x)q=0,
\end{equation}
which is derived from a new symmetry reduction of the well-known AKNS system, where $q=q(x,t)$ is a complex-valued function of the real variables $x$ and $t$, and $*$ denotes complex conjugation. Eq. \eqref {1} is a new integrable system possessing the Lax pair, infinitely many conservation laws and it is solvable by using the inverse scattering transformation (IST). Eq.\eqref {1} is a $PT$-symmetric system, which appears in many fields of physics, such as
nonlinear optics \cite{0,1,2,3}, complex crystal \cite{3.1} and quantum mechanics \cite{3.2,3.3}. The nonlocal NLS equation (1) has attracted much attention of researchers due to its special properties. For example, by using IST method, Ablowitz and Musslimani obtained its breather solution in time. Refs. \cite{7,8} showed that
Eq. \eqref{1} can simultaneously support both bright and dark soliton solutions. The nonsingular localized dark and antidark soliton interactions for the nonlocal defocusing nonlinear Schr\"{o}dinger equation (NLS$^-$) were discussed by using the Darboux transformation method \cite{8.1}.

On the other hand, Ablowitz and Musslimani \cite{9} also investigated an integrable discrete version of the nonlocal NLS equation \eqref{1}
\begin{eqnarray}\label{2}
i q_{n,t}+q_{n+1}+q_{n-1}-2q_n\pm q_n q^{*}_{-n}(q_{n+1}+q_{n-1})=0,
\end{eqnarray}
which is also a discrete $PT$-symmetric model. In Ref. \cite{9},
discrete one-soliton solutions was derived by using IST method. By employing the Hirota's bilinear method, we constructed the N-soliton solution of the integrable nonlocal discrete focusing NLS (NLS$^+$) equation in Ref \cite{10}. Under continuous limit, the discrete soliton yields the one of nonlocal NLS$^+$ equation, which is not periodic in both space and time.

The possible physical application for nonlocal NLS equation in continuous and discrete cases is attracting much attention. In \cite{19} we have shown that the nonlocal NLS equation and its discrete version are gauge equivalent to the corresponding Heisenberg-like equation and modified Heisenberg-like equation and their discrete version, respectively. Recently, Gadzhimuradov and Agalarov further pointed that the nonlocal NLS$^+$ equation \eqref{1} is gauge equivalent to a coupled Landau-Lifshitz (LL) equation. The physical and geometrical aspects of the coupled LL equation are discussed \cite{20}. We should point that, very recently, the physical application for the nonlocal integrable system has a great progress (e.g., see Ref \cite{lou} where an interesting Alice-Boce physics was reported).

As we know, the relation between discrete NLS equation and discrete HF equation was established  \cite{17,18}.
The discrete NLS$^+$ equation and discrete NLS$^-$ equation
\begin{equation}
iq_{n,t}+q_{n+1}+q_{n-1}-2q_{n}\pm|q_n|^2(q_{n+1}+q_{n-1})=0
\end{equation}
are gauge equivalent to the discrete HF equation
\begin{align}
\dot{\textbf{S}}_{n}=
\frac{2\textbf{ S}_{n}\times\textbf{S}_{n-1}}{1+\textbf{S}_{n}\cdot \textbf{S}_{n-1}}-
\frac{2\textbf{S}_{n+1}\times\textbf{S}_{n}}{1+\textbf{S}_{n+1}\cdot \textbf{S}_{n}},
\end{align}
where $S_n=(s_{1n},s_{2n},s_{3n})\in S^2\subset R^3$, and the discrete modified HF equation
\begin{align}
\dot{\textbf{S}}_{n}=
\frac{2\textbf{ S}_{n}\dot{\times}\textbf{S}_{n-1}}{1-\textbf{S}_{n}\cdot \textbf{S}_{n-1}}-
\frac{2\textbf{S}_{n+1}\dot{\times}\textbf{S}_{n}}{1-\textbf{S}_{n+1}\cdot \textbf{S}_{n}},
\end{align}
where $S_n=(s_{1n},s_{2n},s_{3n})\in H^2 \subset R^{2+1}$, and pseudo inner product $\textbf{a} \cdot \textbf{b} $ and pseudo crosse product $\textbf{a} \dot{\times}\textbf{b}$ are defined by $\textbf{a} \cdot \textbf{b}=a_1b_1+a_2b_2-a_3b_3$,
$\textbf{a}\dot{\times}\textbf{b}= (a_2b_3-a_3b_2,a_3b_1-a_1b_3,-(a_1b_2-a_2b_1) )$, respectively.

So, we ask a question:  whether discrete nonlocal NLS equation is gauge equivalent to coupled discrete HF equation or not?
In this paper, we will address the topic. We will show that the nonlocal discrete NLS$^+$ and NLS$^-$ equation \eqref{2} are, respectively, gauge equivalent to the discrete coupled HF equation and the modified discrete coupled HF equation. Under the continuous limit, the discrete coupled HF equation and the modified discrete coupled HF equation yield the coupled HF equation and the modified coupled HF equation. Based on the solution of nonlocal NLS$^-$ equation, the solution of the modified coupled HF equation is constructed. Our results establish the relation of nonlocal NLS equation and coupled HF equation in the discrete and continuous cases.
\section{From discrete nonlocal NLS$^+$ equation to discrete coupled HF equation }
\quad In this section, we will show the relation of discrete nonlocal NLS$^+$ equation and discrete coupled HF equation. Under the continuous limit, the relation yields a fact that nonlocal NLS$^+$ equation is gauge equivalent to coupled HF equation.\\

\textbf{2.1 From discrete nonlocal NLS$^+$ equation to discrete coupled HF equation}\\
The nonlocal discrete NLS$^+$ equation
\begin{equation}\label{Q1}
i q_{n,\tau}+q_{n+1}+q_{n-1}-2q_n+ q_n q^{*}_{-n}(q_{n+1}+q_{n-1})=0
\end{equation}
has the following discrete Lax pair
\begin{equation}\label{Q2}
E\varphi_n=U_n \varphi_n,\quad \varphi_{n,\tau}=V_n \varphi_n,
\end{equation}
with
\begin{align*}
U_n=&\left(
\begin{array}{cc}
z & q^{*}_{-n}z^{-1} \\
- q_n z & z^{-1}  \\
\end{array}\right),\\ \nonumber
V_n=&i \left(
\begin{array}{cc}
1-z^2+z-z^{-1}- q^{*}_{-n}q_{n-1} & -q^{*}_{-n}+q^{*}_{-n+1}z^{-2} \\
- q_n + q_{n-1}z^2 & -1+z^{-2}+z-z^{-1}+ q_n q^{*}_{-n+1}  \\
\end{array}\right).
\end{align*}
Considering discrete gauge transformation
\begin{equation}
\varphi_n=G_n \tilde{\varphi}_n,\quad\tilde{U}_n=G_{n+1}^{-1}U_nG_n, \quad \tilde{V}_n=G_{n}^{-1}V_nG_n-G_{n}^{-1}G_{n,\tau},
\end{equation}
where $G_n$ satisfies the linear problem
\begin{equation}\label{gn1}
G_{n+1}=U_n(1)G_n, \quad G_{n,\tau}=V_n(1)G_n,
\end{equation}
then the spectral problem (7) changes to
\begin{equation}\label{dHlp}
E\tilde{\varphi}_n=\tilde{U}_n \tilde{\varphi}_n,\quad \tilde{\varphi}_{n,\tau}=\tilde{V}_n \tilde{\varphi}_n,
\end{equation}
where
\begin{align}
\tilde{U}_n=&G_{n+1}^{-1}U_nG_n\\\nonumber
=&\frac{z+z^{-1}}{2}G_{n+1}^{-1}U_n(1)G_n+\frac{z-z^{-1}}{2}G_{n}^{-1}\sigma_3G_{n}\\ \nonumber
\triangleq &\frac{z+z^{-1}}{2}I+\frac{z-z^{-1}}{2}S_n, \\ \nonumber
 \tilde{V}_n=&G_{n}^{-1}(V_n-V_n(1))G_n \\ \nonumber
=&i(z-z^{-1})I+i\left(\frac{z^2+z^{-2}}{2}-1\right)G_n^{-1}\left(
\begin{array}{cc}
-1 & q^{*}_{-n+1} \\
q_{n-1} &  1\\
\end{array}\right)G_n\\\nonumber
&+i\frac{z^2-z^{-2}}{2}G_n^{-1}\left(
\begin{array}{cc}
-1 & -q^{*}_{-n+1} \\
q_{n-1} &  -1\\
\end{array}\right)G_n\\
=&i(z-z^{-1})I+i\left(1-\frac{z^2+z^{-2}}{2}\right)\frac{S_n+S_{n-1}}{1+\frac{1}{2}\textrm{tr}(S_nS_{n-1})}
-i\frac{z^2-z^{-2}}{2}\frac{I+S_{n-1}S_n}{1+\frac{1}{2}\textrm{tr}(S_nS_{n-1})}£¬
\end{align}
with  $S_n=G_{n}^{-1}\sigma_3G_{n}$.
The discrete zero curvature equation $\tilde{U}_{n,\tau}=\tilde{V}_{n+1}\tilde{U}_n-\tilde{U}_n\tilde{V}_n$ yields a discrete HF-like equation \cite{19}
\begin{equation}\label{dH}
\frac{dS_n}{d\tau}=\frac{i[S_{n+1},S_n]}{1+\frac{1}{2}\textrm{tr}(S_{n+1}S_{n})}-\frac{i[S_{n},S_{n-1}]}{1+\frac{1}{2}\textrm{tr}(S_nS_{n-1})}.
\end{equation}
The structure of the matrix $U(1)$ and $V(1)$ implies that the solution $G_n$ of Eq. \eqref{gn1} has the form
\begin{equation}\label{eq7}
G_n=\left(
\begin{array}{cc}
f_{1,n} & \omega^*_{-n}g^{*}_{1,1-n} \\
g_{1,n} & \omega^*_{-n}f^{*}_{1,1-n} \\
\end{array}\right),
\end{equation}
where $\omega_n=\prod \limits_{k=-\infty}^{n}\frac{1}{1+q_kq^*_{-k}}$. We emphasize that the matrix $G_n$ possesses the from since if $(f_{1,n},g_{1,n})^T$ is an eigenfunction, then $(\omega^*_{-n}g^{*}_{1,1-n}, \omega^*_{-n}f^{*}_{1,1-n})^T$ is also an eigenfunction. Hence the matrix $S_n$ can be given by
\begin{equation}\label{eq8}
S_n=\frac{1}{\Gamma}\left(
\begin{array}{cc}
\omega^*_{-n}(f_{1,n}f^{*}_{1,1-n}+g_{1,n}g^{*}_{1,1-n}) & 2\omega^{*2}_{-n}f^{*}_{1,1-n}g^{*}_{1,1-n} \\
-2f_{1,n}g_{1,n} & -\omega^*_{-n}(f_{1,n}f^{*}_{1,1-n}+g_{1,n}g^{*}_{1,1-n}) \\
\end{array}\right),
\end{equation}
where $\Gamma=\omega^*_{-n}(f_{1,n}f^{*}_{1,1-n}-g_{1,n}g^{*}_{1,1-n})$. Let $S_n$ be expressed explicitly as
\begin{eqnarray}\label{sn}
S_n=\left(
\begin{array}{cc}
s_{3,n}& s_{1,n}-is_{2,n}\\
 s_{1,n}+is_{2,n} &  -s_{3,n}\\
\end{array}\right),
\end{eqnarray}
where $s_{j,n}(j=1,2,3)$ are complex-valued functions.
By splitting $S_n=M_n+iL_n$, where
\begin{eqnarray*}
M_n=\left(
\begin{array}{cc}
m_{3,n}& m_{1,n}-im_{2,n}\\
m_{1,n}+im_{2,n} &  -m_{3,n}\\
\end{array}\right),\quad
L_n=\left(
\begin{array}{cc}
l_{3,n}& l_{1,n}-il_{2,n}\\
l_{1,n}+il_{2,n} &  -l_{3,n}\\
\end{array}\right),
\end{eqnarray*}
and noting that $1+\frac{1}{2}\textrm{tr}(S_{n+1}S_{n})=1+\textbf{S}_{n+1}\cdot\textbf{S}_{n}$, then Eq.\eqref{dH} is rewritten in the vector form
\begin{subequations}
\begin{eqnarray}\begin{aligned}\label{cHF1}
\textbf{m}_{n,\tau}=&\frac{2\Delta_1(\textbf{m}_n\times\textbf{m}_{n+1}-\textbf{l}_n\times\textbf{l}_{n+1})
-2\Delta_2(\textbf{m}_n\times\textbf{l}_{n+1}-\textbf{m}_{n+1}\times\textbf{l}_n)}{\Delta_1^2+\Delta_2^2}\\
&+\frac{2\Delta_3(\textbf{m}_n\times\textbf{m}_{n-1}-\textbf{l}_n\times\textbf{l}_{n-1})
-2\Delta_4(\textbf{m}_n\times\textbf{l}_{n-1}-\textbf{m}_{n-1}\times\textbf{l}_n)}{\Delta_3^2+\Delta_4^2},
\end{aligned}
\end{eqnarray}
\begin{eqnarray}\begin{aligned}\label{cHF2}
\textbf{l}_{n,\tau}=&\frac{2\Delta_2(\textbf{m}_n\times\textbf{m}_{n+1}-\textbf{l}_n\times\textbf{l}_{n+1})
+2\Delta_1(\textbf{m}_n\times\textbf{l}_{n+1}-\textbf{m}_{n+1}\times\textbf{l}_n)}{\Delta_1^2+\Delta_2^2}\\
&+\frac{2\Delta_4(\textbf{m}_n\times\textbf{m}_{n-1}-\textbf{l}_n\times\textbf{l}_{n-1})
+2\Delta_3(\textbf{m}_n\times\textbf{l}_{n-1}-\textbf{m}_{n-1}\times\textbf{l}_n)}{\Delta_3^2+\Delta_4^2},
\end{aligned}
\end{eqnarray}
\end{subequations}
where real-valued vectors $\textbf{m}_n=(m_{1,n},m_{2,n},m_{3,n}),\textbf{l}_n=(l_{1,n},l_{2,n},l_{3,n})$ satisfy
\begin{eqnarray}\begin{aligned}
\textbf{m}_n\cdot\textbf{m}_n-\textbf{l}_n\cdot\textbf{l}_n=1, \quad \textbf{m}_n\cdot\textbf{l}_n=0,
\end{aligned}
\end{eqnarray}
and
\begin{eqnarray}\begin{aligned}
\Delta_1=1+\textbf{m}_n\cdot\textbf{m}_{n+1}-\textbf{l}_n\cdot\textbf{l}_{n+1},\quad \Delta_2=\textbf{m}_n\cdot\textbf{l}_{n+1}+\textbf{l}_n\cdot\textbf{m}_{n+1},\\
\Delta_3=1+\textbf{m}_n\cdot\textbf{m}_{n-1}-\textbf{l}_n\cdot\textbf{l}_{n-1},\quad \Delta_4=\textbf{m}_n\cdot\textbf{l}_{n-1}+\textbf{l}_n\cdot\textbf{m}_{n-1},\\
1+\frac{1}{2}\textrm{tr}(S_{n+1}S_{n})=\Delta_1+i \Delta_2,\quad 1+\frac{1}{2}\textrm{tr}(S_nS_{n-1})=\Delta_3+i \Delta_4.
\end{aligned}
\end{eqnarray}
The relation between $m_{j,n},l_{j,n}$ and $s_{j,n}$ is $m_{j,n}=\textrm{Re}(s_{j,n}), l_{j,n}=\textrm{Im}(s_{j,n})(j=1,2,3).$ Here, the inner product and the cross product are defined in $R^3$. We thus have shown that the nonlocal discrete NLS$^+$ equation \eqref{Q1} is gauge equivalent to the discrete coupled HF equation (17).
Note that under the spatial parity symmetry $q(n,\tau)=q(-n,\tau)$, the discrete nonlocal NLS$^+$ equation becomes the classical discrete NLS$^+$ equation. The matrix $S_n$ \eqref{sn} is transformed into the Hermitian and $\textbf{l}_n=0$, thus the discrete coupled HF equation (17) reduces to discrete HF model
\begin{eqnarray}\begin{aligned}
\textbf{m}_{n,\tau}=&\frac{2\textbf{m}_n\times\textbf{m}_{n+1}
}{1+\textbf{m}_n\cdot\textbf{m}_{n+1}}+\frac{2\textbf{m}_n\times\textbf{m}_{n-1}}
{1+\textbf{m}_n\cdot\textbf{m}_{n-1}}.
\end{aligned}
\end{eqnarray}
\textbf{2.2 From nonlocal NLS$^+$ equation to coupled HF equation through continuous limit}\\
In this subsection we will show how to get coupled HF equation from nonlocal NLS$^+$ equation through continuous limit. Let $z= 1+i\lambda \epsilon, q_{n}(\tau)=\epsilon q(x,t), x= n \epsilon, t=\epsilon^2 \tau$ and $\varphi_n\thicksim \varphi$,  the continuous limit of the discrete Lax pair \eqref{Q2} gives
\begin{equation}
\varphi_x=U \varphi,\quad \varphi_{t}=V\varphi,
\end{equation}
where
\begin{align*}
U=\left(
\begin{array}{cc}
i\lambda & q^{*}(-x) \\
 -q  & -i\lambda \\
\end{array}\right), \quad
V= \left(
\begin{array}{cc}
2i\lambda^2-iq q^{*}(-x) & 2\lambda q^{*}(-x)-iq^{*}_{x}(-x) \\
-2\lambda q-iq_x & -2i\lambda^2+iq q^{*}(-x) \\
\end{array}\right),
\end{align*}
with $q^{*}_x(-x) =-(q^{*}(-x))_x $. Here we have used the Taylor expansion
$$q_{n\pm 1}\thicksim \epsilon (q \pm \epsilon q_x +\frac{\epsilon^2}{2}q_{xx}+\cdot\cdot\cdot),\quad
q^*_{-n\pm 1}\thicksim \epsilon (q^*(-x) \pm \epsilon q^*_x(-x) +\frac{\epsilon^2}{2}q^*_{xx}(-x)+\cdot\cdot\cdot).$$
The integrability condition $U_t-V_x+[U,V]=0$ yields the nonlocal NLS$^+$ equation $i q_t+q_{xx}+ 2qq^{*}(-x)q=0$. After the substitution
\begin{equation}\label{tihuan}
z\rightarrow 1+i\lambda \epsilon, \quad x\rightarrow n \epsilon,\quad t\rightarrow\epsilon^2 \tau, \quad S_n \rightarrow S, \quad \tilde{\varphi}_n\rightarrow \tilde{\varphi},
\end{equation}
the continuous limit of the spectral problem \eqref{dHlp} leads to
\begin{equation}\label{dHlp1}
\tilde{\varphi}_x=\tilde{U} \tilde{\varphi},\quad \tilde{\varphi}_{t}=\tilde{V} \tilde{\varphi},
\end{equation}
where
\begin{equation}
\tilde{U}=i \lambda S,\quad \tilde{V} = \lambda S S_x+2i\lambda^2 S, \qquad  S=G^{-1}\sigma_3 G.
\end{equation}
The integrability condition of \eqref{dHlp1} leads to
\begin{eqnarray}
S_t=-\frac{i}{2}[S,S_{xx}],
\end{eqnarray}
which is just HF model in the matrix form (see Refs. \cite{19, 20}).
Moreover, one can check that, under the substitution $M_n\rightarrow M, L_n\rightarrow L, x\rightarrow n \epsilon,t\rightarrow\epsilon^2 \tau$ as $\epsilon \rightarrow 0$, the discrete coupled HF equation (17) yields the following coupled HF equation \cite {20}
\begin{eqnarray}\begin{aligned}\label{cnLL}
\textbf{m}_t&=\textbf{m}\times\textbf{m}_{xx}-\textbf{l}\times\textbf{l}_{xx},\\
\textbf{l}_t&=\textbf{m}\times\textbf{l}_{xx}+\textbf{l}\times\textbf{m}_{xx},
\end{aligned}
\end{eqnarray}
where real-valued vectors $\textbf{m}=(m_1,m_2,m_3),\textbf{l}=(l_1,l_2,l_3)$ satisfy
$\textbf{m}^2-\textbf{l}^2=1,\quad \textbf{m}\cdot\textbf{l}=0.$  We remark here that under the spatial parity symmetry $q(x,t)=q(-x,t)$, the nonlocal NLS$^+$ equation becomes the classical NLS$^+$ equation. In this case, the matrix $S$ is transformed into the Hermitian and $\textbf{l}=0$. Thus the coupled HF system \eqref{cnLL} reduces to the HF equation,
\begin{equation}\label{4}
\textbf{m}_t=\textbf{m}\times \textbf{m}_{xx},
\end{equation}
where $\textbf{m}=(m_1,m_2,m_3)\in S^2$ in $R^3$. This is consistent with a well-known fact that the NLS$^+$ equation
is gauge equivalent to the HF equation \cite{13, 14,16}.

\section{From discrete nonlocal NLS$^-$ equation to coupled discrete modified HF equation }
\quad In this section, we will show that the nonlocal discrete NLS$^-$ equation is gauge equivalent to a modified discrete coupled HF equation.
Under the continuous limit, the modified discrete coupled HF equation yields a modified coupled HF equation. This implies that the nonlocal NLS$^-$ equation is gauge equivalent
to the modified coupled HF equation. The exact solution of the modified coupled HF equations is constructed through the soliton solution of nonlocal NLS$^-$ equation.

\textbf{3.1 From discrete nonlocal NLS$^-$ equation to coupled discrete modified HF equation}\\
For the nonlocal discrete NLS$^-$ equation
\begin{equation}\label{fu1}
i q_{n,\tau}+ q_{n+1}+q_{n-1}-2q_n- q_n q^{*}_{-n}(q_{n+1}+q_{n-1})=0,
\end{equation}
its discrete Lax pair is
\begin{equation}\label{fu2}
E\varphi_n=U_n \varphi_n,\quad \varphi_{n,\tau}=V_n \varphi_n,
\end{equation}
with
\begin{align*}
U_n=&\left(
\begin{array}{cc}
z & q^{*}_{-n}z^{-1} \\
q_n z & z^{-1}  \\
\end{array}\right),\\ \nonumber
V_n=&i \left(
\begin{array}{cc}
1-z^2+z-z^{-1}+q^{*}_{-n}q_{n-1} & -q^{*}_{-n}+q^{*}_{-n+1}z^{-2} \\
 q_n - q_{n-1}z^2 & -1+z^{-2}+z-z^{-1}- q_n q^{*}_{-n+1}  \\
\end{array}\right).
\end{align*}
Set $S_n=-iG_{n}^{-1}\sigma_3G_{n}$, where $G_n$ satisfies the linear problem
\begin{equation}\label{gn2}
G_{n+1}=U_n(1)G_n, \quad G_{n,\tau}=V_n(1)G_n.
\end{equation}
Under discrete gauge transformation
\begin{equation}
\varphi_n=G_n\tilde{\varphi}_n, \quad \tilde{U}_n=G_{n+1}^{-1}U_nG_n, \quad \tilde{V}_n=G_{n}^{-1}V_nG_n-G_{n}^{-1}G_{n,\tau},
\end{equation}
the spectral problem (29) changes to
\begin{equation}\label{nLp}
E\tilde{\varphi}_n=\tilde{U}_n \tilde{\varphi}_n,\quad \tilde{\varphi}_{n,\tau}=\tilde{V}_n \tilde{\varphi}_n,
\end{equation}
where
\begin{align}\label{17a}
\tilde{U}_n=&-iG_{n}^{-1}U_n^{-1}(1)U_nG_n=\frac{z+z^{-1}}{2}I+i\frac{z-z^{-1}}{2}S_n, \\\nonumber
 \tilde{V}_n=&G_{n}^{-1}(V_n-V_n(1))G_n \\ \nonumber
=&i(z-z^{-1})I+i\left(\frac{z^2+z^{-2}}{2}-1\right)G_n^{-1}\left(
\begin{array}{cc}
-1 & q^{*}_{-n-1} \\
-q_{n-1} &  1\\
\end{array}\right)G_n\\\nonumber
&+i\frac{z^2-z^{-2}}{2}G_n^{-1}\left(
\begin{array}{cc}
-1 & -q^{*}_{-n+1} \\
-q_{n-1} &  -1\\
\end{array}\right)G_n\\ \label{17b}
=&i(z-z^{-1})I+\frac{z^2+z^{-2}-2}{2}\frac{S_n+S_{n-1}}{1-\frac{1}{2}\textrm{tr}(S_nS_{n-1})}
-i\frac{z^2-z^{-2}}{2}\frac{I-S_{n-1}S_n}{1-\frac{1}{2}\textrm{tr}(S_nS_{n-1})}.
\end{align}
The discrete zero curvature equation $\tilde{U}_{n,\tau}=\tilde{V}_{n+1}\tilde{U}_n-\tilde{U}_n\tilde{V}_n$ yields a discrete modified HF-like model \cite{19}
\begin{equation}\label{dmH2}
\frac{dS_n}{d\tau}=\frac{[S_{n},S_{n-1}]}{1-\frac{1}{2}\textrm{tr}(S_nS_{n-1})}-\frac{[S_{n+1},S_n]}{1-\frac{1}{2}\textrm{tr}(S_{n+1}S_{n})}.
\end{equation}
The structure of the matrix $U(1)$ and $V(1)$ implies that the solution $G_n$ of Eq. \eqref{gn2} has the form
\begin{equation}\label{eqgn}
G_n=\left(
\begin{array}{cc}
f_{1,n} & \omega^*_{-n}g^{*}_{1,1-n} \\
g_{1,n} & -\omega^*_{-n}f^{*}_{1,1-n} \\
\end{array}\right),
\end{equation}
where $\omega_n=\prod \limits_{k=-\infty}^{n}\frac{1}{1-q_kq^*_{-k}}$. We should remark here that if $(f_{1,n},g_{1,n})^T$ is an eigenfunction, then $(\omega^*_{-n}g^{*}_{1,1-n}, -\omega^*_{-n}f^{*}_{1,1-n})^T$ is also an eigenfunction. This means that the matrix $G_n$ possesses the from.  Hence the matrix $S_n$ can be written as
\begin{equation}\label{sn2}
S_n=\frac{1}{\Theta}\left(
\begin{array}{cc}
i\omega^*_{-n}(g_{1,n}g^{*}_{1,1-n}-f_{1,n}f^{*}_{1,1-n}) & -2i\omega^{*2}_{-n}f^{*}_{1,1-n}g^{*}_{1,1-n} \\
-2if_{1,n}g_{1,n} & i\omega^*_{-n}(f_{1,n}f^{*}_{1,1-n}-g_{1,n}g^{*}_{1,1-n}) \\
\end{array}\right),
\end{equation}
where $\Theta=\omega^*_{-n}(f_{1,n}f^{*}_{1,1-n}+g_{1,n}g^{*}_{1,1-n})$.
Let $S_n$ be expressed explicitly as
\begin{eqnarray}\label{sn1}
S_n=\left(
\begin{array}{cc}
s_{1,n}& i(s_{3,n}-s_{2,n})\\
i( s_{3,n}+s_{2,n}) &  -s_{1,n}\\
\end{array}\right),
\end{eqnarray}
where $s_{j,n}(j=1,2,3)$ are complex-valued functions.
By splitting $S_n=M_n+iL_n$, where
\begin{eqnarray*}
M_n=\left(
\begin{array}{cc}
m_{1,n}& i(m_{3,n}-m_{2,n})\\
i(m_{3,n}+m_{2,n}) &  -m_{1,n}\\
\end{array}\right),\quad
L_n=\left(
\begin{array}{cc}
l_{1,n}& i(l_{3,n}-l_{2,n})\\
i(l_{3,n}+l_{2,n}) &  -l_{1,n}\\
\end{array}\right),
\end{eqnarray*}
then Eq.\eqref{dmH2} is rewritten in the following vector form:
\begin{subequations}\label{dcmFH}
\begin{eqnarray}\begin{aligned}\label{cmHF1}
\textbf{m}_{n,\tau}=&\frac{2\Omega_1(\textbf{m}_n\dot{\times}\textbf{m}_{n-1}-\textbf{l}_n\dot{\times}\textbf{l}_{n-1})
-2\Omega_2(\textbf{m}_n\dot{\times}\textbf{l}_{n-1}-\textbf{m}_{n-1}\dot{\times}\textbf{l}_n)}{\Omega_1^2+\Omega_2^2}\\
-&\frac{2\Omega_3(\textbf{m}_n\dot{\times}\textbf{m}_{n+1}-\textbf{l}_n\dot{\times}\textbf{l}_{n+1})
-2\Omega_4(\textbf{m}_n\dot{\times}\textbf{l}_{n+1}-\textbf{m}_{n+1}\dot{\times}\textbf{l}_n)}{\Omega_3^2+\Omega_4^2},\\
\end{aligned}
\end{eqnarray}
\begin{eqnarray}\begin{aligned}\label{cmHF2}
\textbf{l}_{n,\tau}=&\frac{2\Omega_2(\textbf{m}_n\dot{\times}\textbf{m}_{n-1}-\textbf{l}_n\dot{\times}\textbf{l}_{n-1})
+2\Omega_1(\textbf{m}_n\dot{\times}\textbf{l}_{n-1}-\textbf{m}_{n-1}\dot{\times}\textbf{l}_n)}{\Omega_1^2+\Omega_2^2}\\
+&\frac{2\Omega_4(\textbf{m}_n\dot{\times}\textbf{m}_{n+1}-\textbf{l}_n\dot{\times}\textbf{l}_{n+1})
+2\Omega_3(\textbf{m}_n\dot{\times}\textbf{l}_{n+1}-\textbf{m}_{n+1}\dot{\times}\textbf{l}_n)}{\Omega_3^2+\Omega_4^2},\\
\end{aligned}
\end{eqnarray}
\end{subequations}
where real-valued vectors $\textbf{m}_n=(m_{1,n},m_{2,n},m_{3,n}),\textbf{l}_n=(l_{1,n},l_{2,n},l_{3,n})$ satisfy
\begin{eqnarray}\begin{aligned}
\textbf{m}_n\cdot\textbf{m}_n-\textbf{l}_n\cdot\textbf{l}_n=-1, \quad \textbf{m}_n\cdot\textbf{l}_n=0,
\end{aligned}
\end{eqnarray}
and
\begin{eqnarray}\begin{aligned}
&\Omega_1=1-\textbf{m}_n\cdot\textbf{m}_{n-1}+\textbf{l}_n\cdot\textbf{l}_{n-1},\quad
\Omega_2=\textbf{m}_n\cdot\textbf{l}_{n-1}+\textbf{l}_n\cdot\textbf{m}_{n-1},\\
&\Omega_3=1-\textbf{m}_n\cdot\textbf{m}_{n+1}+\textbf{l}_n\cdot\textbf{l}_{n+1},\quad
\Omega_4=\textbf{m}_n\cdot\textbf{l}_{n+1}+\textbf{l}_n\cdot\textbf{m}_{n+1},\\
&1-\frac{1}{2}\textrm{tr}(S_{n}S_{n-1})=\Omega_1+i \Omega_2,\quad 1-\frac{1}{2}\textrm{tr}(S_{n+1}S_n)=\Omega_3+i \Omega_4.
\end{aligned}
\end{eqnarray}
The relation between $m_{j,n},l_{j,n}$ and $s_{j,n}$ is $m_{j,n}=\textrm{Re}(s_{j,n}), l_{j,n}=\textrm{Im}(s_{j,n})(j=1,2,3).$ In the above equations, the pseudo inner product
$\cdot$ and the pseudo cross product $\dot{\times}$ are defined by the vector in $R^{2+1}$. Hence, the nonlocal discrete NLS$^-$ equation \eqref{fu1} is gauge equivalent to the modified discrete coupled HF equation \eqref{dcmFH}.

Note that under the spatial parity symmetry $q(n,\tau)=q(-n,\tau)$, the discrete nonlocal NLS$^-$ equation becomes the classical discrete NLS$^-$ equation. In this case, the matrix $S_n$ \eqref{sn1} is transformed into the Hermitian and $\textbf{l}_n=0$. Thus the discrete coupled modified HF system \eqref{dcmFH} reduces to the modified discrete HF model
\begin{eqnarray}\begin{aligned}
\textbf{m}_{n,\tau}=&\frac{2\textbf{m}_n\dot{\times}\textbf{m}_{n-1}
}{1-\textbf{m}_n\cdot\textbf{m}_{n-1}}-\frac{2\textbf{m}_{n+1}\dot{\times}\textbf{m}_n}
{1-\textbf{m}_{n+1}\cdot\textbf{m}_n}.
\end{aligned}
\end{eqnarray}

\textbf{3.2  From nonlocal NLS$^-$ equation to coupled modified HF equation through continuous limit }\\
Considering the transformation
 $$z= 1+\lambda \epsilon, q_{n}(\tau)=\epsilon q(x,t), x= n \epsilon, t=\epsilon^2 \tau, \varphi_n\thicksim \varphi,$$
the continuous limit of the discrete Lax pair \eqref{fu2} gives
 \begin{equation}
\varphi_x=U \varphi,\quad \varphi_t=V \varphi
\end{equation}
where
\begin{align*}
U=\left(
\begin{array}{cc}
\lambda & q^{*}(-x) \\
 q & -\lambda  \\
\end{array}\right),\quad
V=i \left(
\begin{array}{cc}
-2\lambda^2+ qq^{*}(-x) & -2\lambda q^{*}(-x)+q^{*}_x(-x) \\
- 2\lambda q + q_x & 2\lambda^2- qq^{*}(-x)  \\
\end{array}\right).
\end{align*}
The integrability condition $U_t-V_x+[U,V]=0$ yields the nonlocal NLS$^-$ equation $i q_t+q_{xx}-2qq^{*}(-x)q=0$.
Let $G_n\rightarrow G$, the linear problem \eqref{gn2} changes to
\begin{equation}\label{mgt}
G_{x}=U(0)G, \quad G_{t}=V(0)G.
\end{equation}
Similarly, through the substitution $z\rightarrow 1+\lambda \epsilon,  x\rightarrow n \epsilon, t\rightarrow\epsilon^2 \tau,  S_n \rightarrow S, \tilde{\varphi}_n\rightarrow \tilde{\varphi}$, the continuous limit of the spectral problem \eqref{nLp} yields
\begin{equation}
\tilde{\varphi}_x=\tilde{U} \tilde{\varphi},\quad \tilde{\varphi}_{t}=\tilde{V} \tilde{\varphi},
\end{equation}
where
\begin{align}
\tilde{U}=\lambda G^{-1}p_1 G \triangleq i\lambda S, \quad
\tilde{V}=2\lambda^2S+i \lambda SS_x,
\end{align}
The compatibility condition $\tilde{U}_t-\tilde{V}_x+[\tilde{U},\tilde{V}]=0$ yields a modified HF-like equation in the matrix form \cite{19},
\begin{eqnarray}\label{mHE}
S_t=\frac{1}{2}[S,S_{xx}].
\end{eqnarray}
Moreover, $S_n=M_n+iL_n\rightarrow S=M+iL$, it is direct to verify that under continuous limit, the discrete modified coupled HF system \eqref{dcmFH}
 leads to a modified coupled HF equation
\begin{eqnarray}\begin{aligned}\label{eq4.22}
\textbf{m}_t&=\textbf{m}\dot{\times}\textbf{m}_{xx}-\textbf{l}\dot{\times}\textbf{l}_{xx},\\
\textbf{l}_t&=\textbf{m}\dot{\times}\textbf{l}_{xx}+\textbf{l}\dot{\times}\textbf{m}_{xx},
\end{aligned}
\end{eqnarray}
where real-valued vectors $\textbf{m}=(m_1,m_2,m_3),\textbf{l}=(l_1,l_2,l_3)$ satisfy
\begin{eqnarray}\begin{aligned}
\textbf{l}^2-\textbf{m}^2=1,\quad \textbf{m}\textbf{l}=0.
\end{aligned}
\end{eqnarray}
The relation between $m_j,l_j$ and $s_j$ is $m_j=\textrm{Re}(s_j), l_j=\textrm{Im}(s_j)(j=1,2,3).$
Here $M= m_1p_1+i m_2 p_2+im_3p_3,L= l_1p_1+i l_2 p_2+il_3p_3$, where $\pmb{p}=(p_1,p_2,p_3)$ with
\begin{eqnarray*}
p_1=\left(
\begin{array}{cc}
1 & 0\\
0 & -1\\
\end{array}\right), \quad p_2=\left(
\begin{array}{cc}
0 & -1\\
1 & 0\\
\end{array}\right),\quad p_3=\left(
\begin{array}{cc}
0 & 1\\
1 & 0\\
\end{array}\right).
\end{eqnarray*}
We remark here that under the spatial parity symmetry $q(x,t)=q(-x,t)$, the nonlocal NLS$^-$ equation becomes the classical NLS$^-$ equation. The matrix $S$ is transformed into the Hermitian and $\textbf{l}=0$. Thus the coupled modified HF system \eqref{eq4.22} reduces to the modified HF equation \begin{eqnarray}\begin{aligned}
\textbf{m}_t=\textbf{m}\dot{\times}\textbf{m}_{xx}.
\end{aligned}
\end{eqnarray}

\textbf{3.3 Solution of coupled modified HF equation}

From the structure of the matrix $U(0)$ and $V(0)$, we can see that the matrix $G$ in \eqref{mgt} has the form
\begin{equation*}
G=\left(
\begin{array}{cc}
f & -g^{*}(-x) \\
g & f^{*}(-x) \\
\end{array}\right).
\end{equation*}
This means that if $(f,g)^T$ is an eigenfunction, then $( -g^{*}(-x),f^{*}(-x))^T$ is also an eigenfunction. Hence the matrix $S$ is given by
\begin{equation}\label{s2}
S=-iG^{-1}p_1 G=i\Omega\left(
\begin{array}{cc}
gg^{*}(-x)-ff^{*}(-x) & 2f^{*}(-x)g^{*}(-x) \\
2fg & ff^{*}(-x)-gg^{*}(-x) \\
\end{array}\right),
\end{equation}
where $\Omega=\left(ff^{*}(-x)+gg^{*}(-x)\right)^{-1}$.
It is obvious that matrix $S$ is not Hermitian. It can be written as
\begin{eqnarray*}
S=\left(
\begin{array}{cc}
s_{1}& i(s_{3}-s_{2})\\
 i(s_{3}+s_{2}) &  -s_{1}\\
\end{array}\right),
\end{eqnarray*}
where the components $s_j=s_j(x,t)$ of complex-valued vector $(s_1,s_2,s_3)$ are given by
\begin{equation}\begin{aligned}\label{eq4.23}
s_1&=i\Omega \left(gg^{*}(-x)-ff^{*}(-x)\right),\quad
s_2=\Omega \left(fg-f^{*}(-x)g^{*}(-x)\right),\\
s_3&=\Omega \left(fg+f^{*}(-x)g^{*}(-x)\right).
\end{aligned}\end{equation}
It is direct to check that $s_j$ satisfies
\begin{eqnarray*}
s_1(x,t)=-s_1^{*}(-x,t),\quad s_2(x,t)=-s_2^{*}(-x,t),\quad s_3(x,t)=s_3^{*}(-x,t).
\end{eqnarray*}

Let us construct the solution of the modified coupled HF Eq. \eqref{eq4.22} based on the solution $q(x,t)$ of nonlocal defocusing NLS equation given by
\begin{align}\label{fuone}
q(x,t)=-2ibe^{-2ax+4i(a^2-b^2)t}\sech(8abt+2ibx),\quad b\neq 0.
\end{align}
Solving Eq. \eqref{mgt} obtains
\begin{equation*}
G=\left(
\begin{array}{cc}
 \frac{-ia+ b\tanh(8abt+2ibx)}{\sqrt{a^2+b^2}}& -\frac{b e^{2ax-4i(a^2-b^2)t}\sech(8abt+2ibx)}{\sqrt{a^2+b^2}}  \\
\frac{b e^{-2ax+4i(a^2-b^2)t}\sech(8abt+2ibx)}{\sqrt{a^2+b^2}} & \frac{ia+b\tanh(8abt+2ibx)}{\sqrt{a^2+b^2}} \\
\end{array}\right).
\end{equation*}
Then by using the expression $S=-iG^{-1}p_1 G$, we have
\begin{eqnarray*}
S=\left(
\begin{array}{cc}
 -i+\frac{2ib^2\sech^2(8abt+2ibx)}{a^2+b^2}& -\frac{2b e^{-4i(a^2-b^2)t+2ax}\sech(8abt+2ibx)\left(a-ib \tanh(8abt+2ibx)\right)}{a^2+b^2}  \\
\frac{2b e^{4i(a^2-b^2)t-2ax}\sech(8abt+2ibx)\left(a+ib \tanh(8abt+2ibx)\right)}{a^2+b^2} & i-\frac{2ib^2\sech^2(8abt+2ibx)}{a^2+b^2} \\
\end{array}\right).
\end{eqnarray*}
i.e.
\begin{eqnarray*}
\begin{aligned}
s_1&= -i+\frac{2ib^2\sech^2(8abt+2ibx)}{a^2+b^2},\\
s_2&=\frac{2ib\sech(8abt+2ibx)\left(-a\cos(4(a^2-b^2)t+2iax)+b \sin(4(a^2-b^2)t+2iax)\tanh(8abt+2ibx)\right)}{a^2+b^2},\\
s_3&=\frac{2b\sech(8abt+2ibx)\left(a\sin(4(a^2-b^2)t+2iax)+b \cos(4(a^2-b^2)t+2iax)\tanh(8abt+2ibx)\right)}{a^2+b^2}.
\end{aligned}
\end{eqnarray*}
Therefore the one-soliton solution of the modified coupled HF equation \eqref{eq4.22} is
\begin{eqnarray}\label{ml}
\begin{aligned}
m_1&=\textrm{Re}(s_1)=\frac{4b^2\sin(4bx)\sinh(16abt)}{(a^2+b^2)(\cos(4bx)+\cosh(16abt))^2},\\
l_1&=\textrm{Im}(s_1)=-1+8b^2\frac{\cos^2(2bx)\cosh^2(8abt)-\sin^2(2bx)\sinh^2(8abt)}{(a^2+b^2)(\cos(4bx)+\cosh(16abt))^2},\\
\end{aligned}
\end{eqnarray}
while the expressions $m_j=\textrm{Re}(s_j)$ and $l_j=\textrm{Im}(s_j)(j=2,3)$ are omitted here because they are too long and complicated.
The singular point of the one-soliton solution occurs at $(x,t)=(\frac{2k+1}{4b}\pi,0),\quad k\in Z$.
The dynamic of the one-soliton solution \eqref{ml} through numerical simulation is shown in the Figs 1 and 2.
\begin{figure*}[t!]
\centering
\subfigure[] {\includegraphics[width=0.25\textwidth]{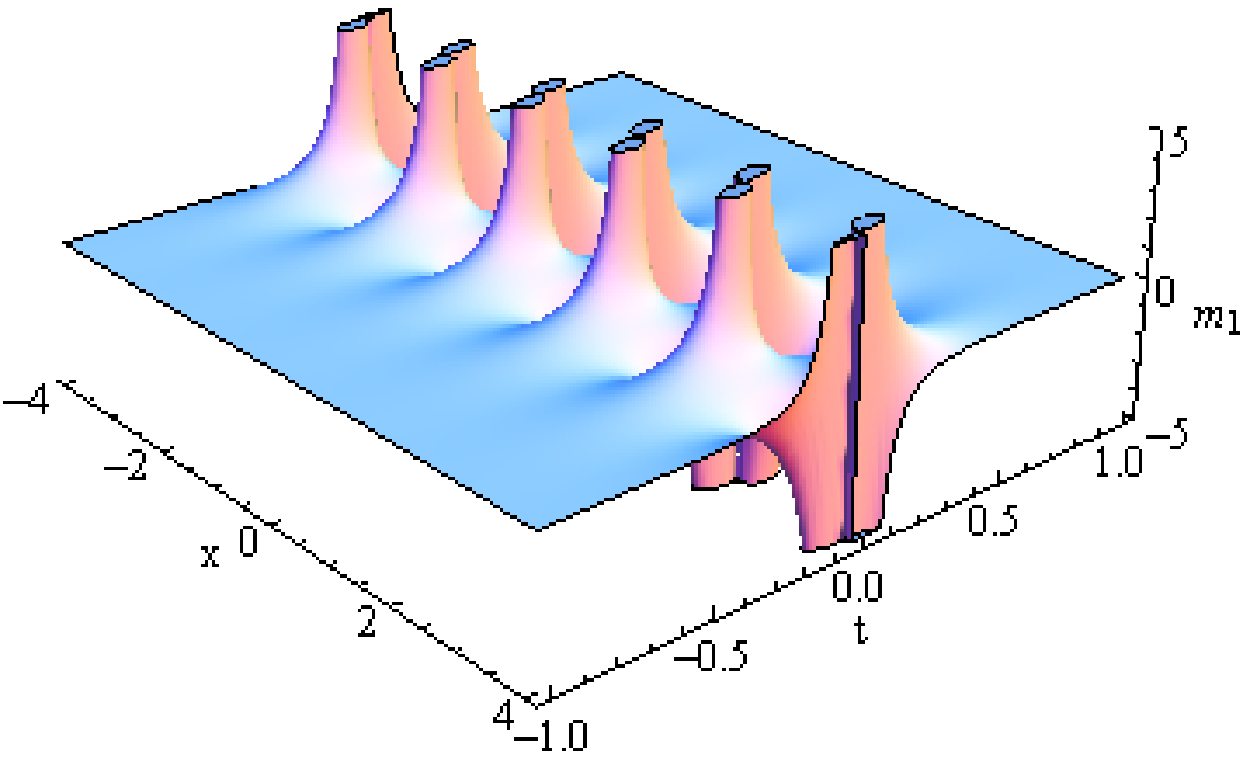}}\quad
\subfigure[] {\includegraphics[width=0.25\textwidth]{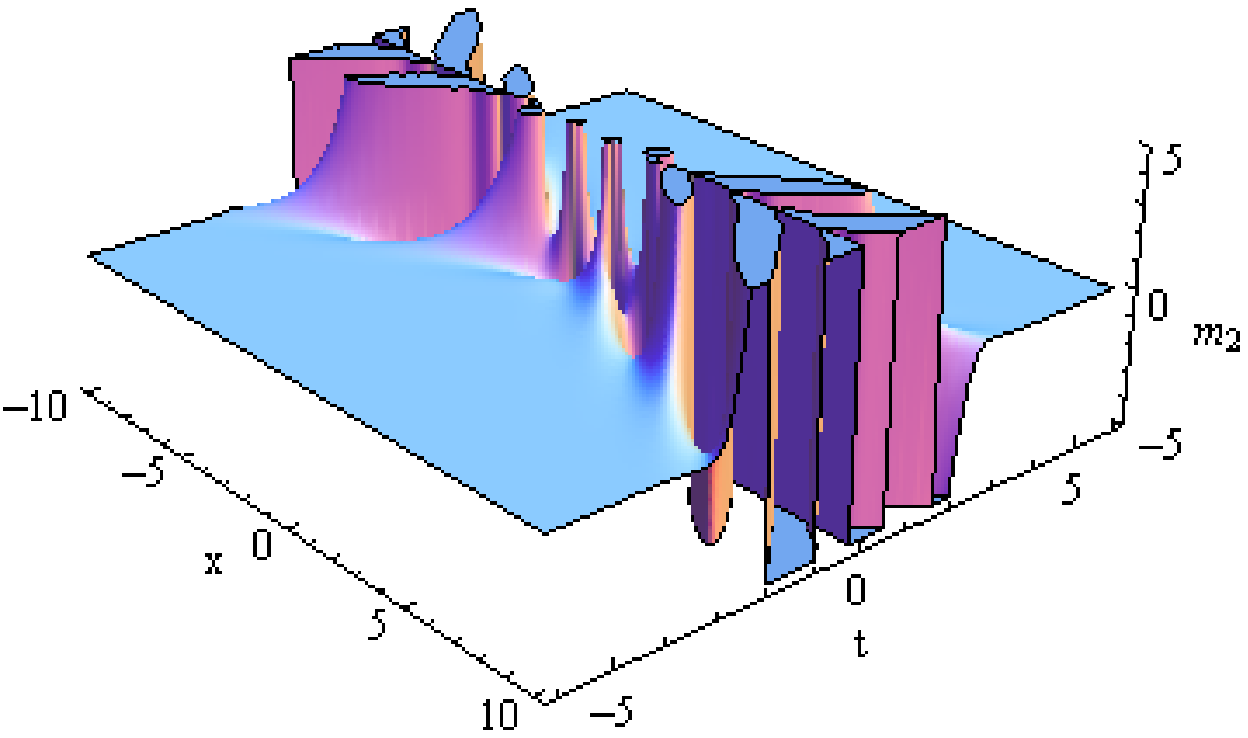}}\quad
\subfigure[] {\includegraphics[width=0.25\textwidth]{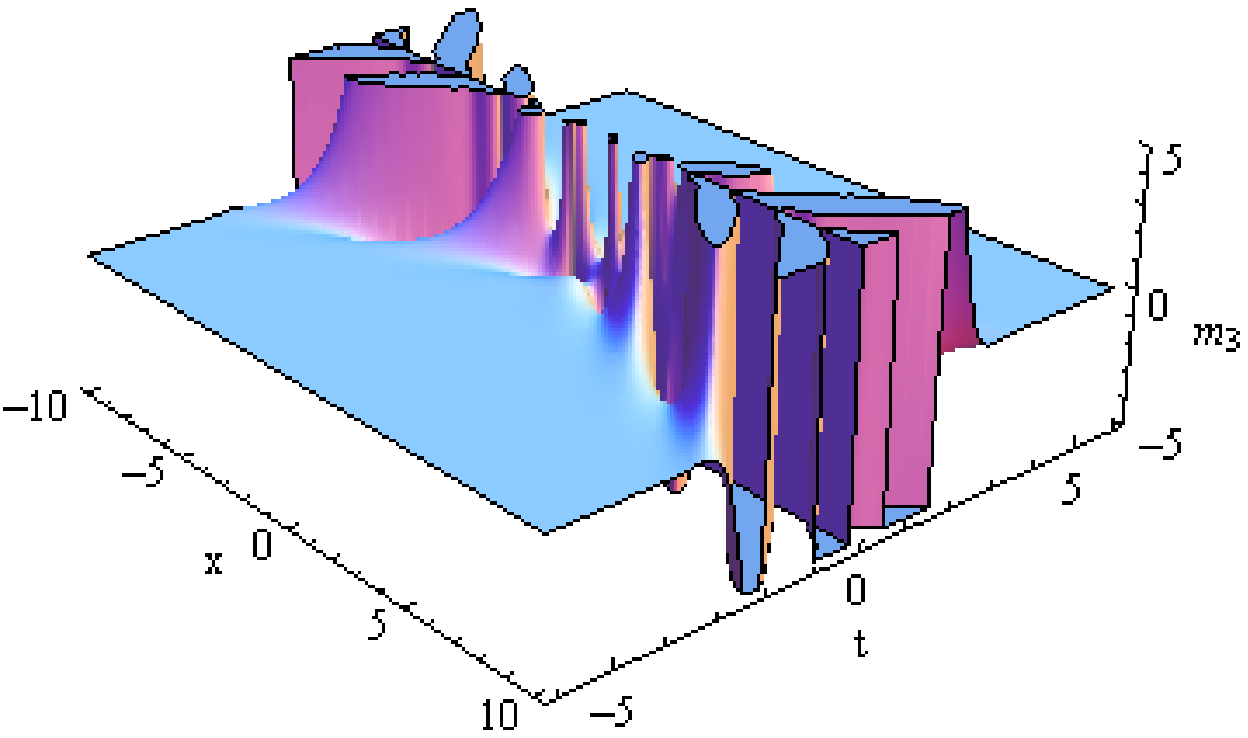}}\\
\subfigure[] {\includegraphics[width=0.23\textwidth]{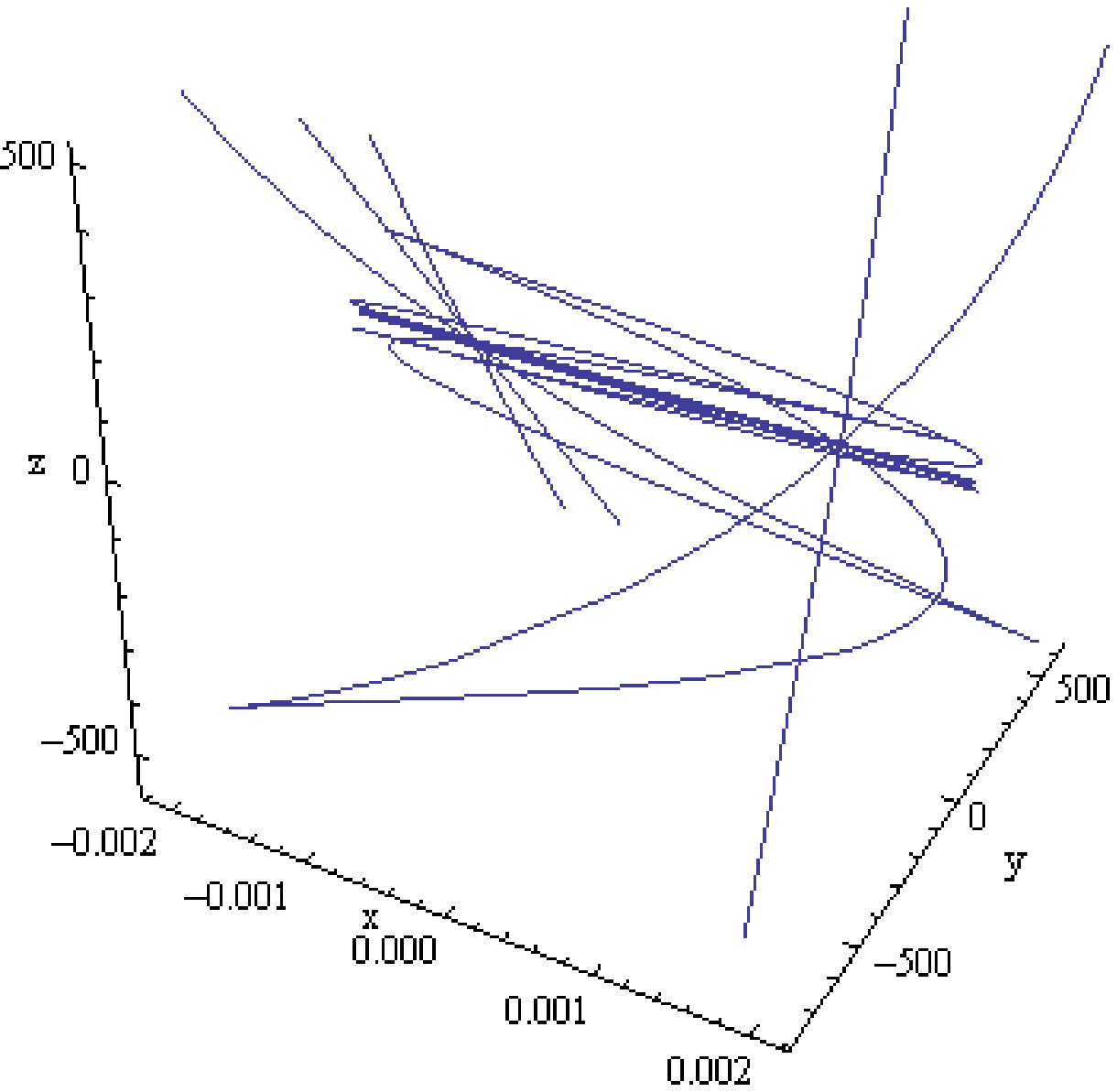}}\quad
\subfigure[] {\includegraphics[width=0.23\textwidth]{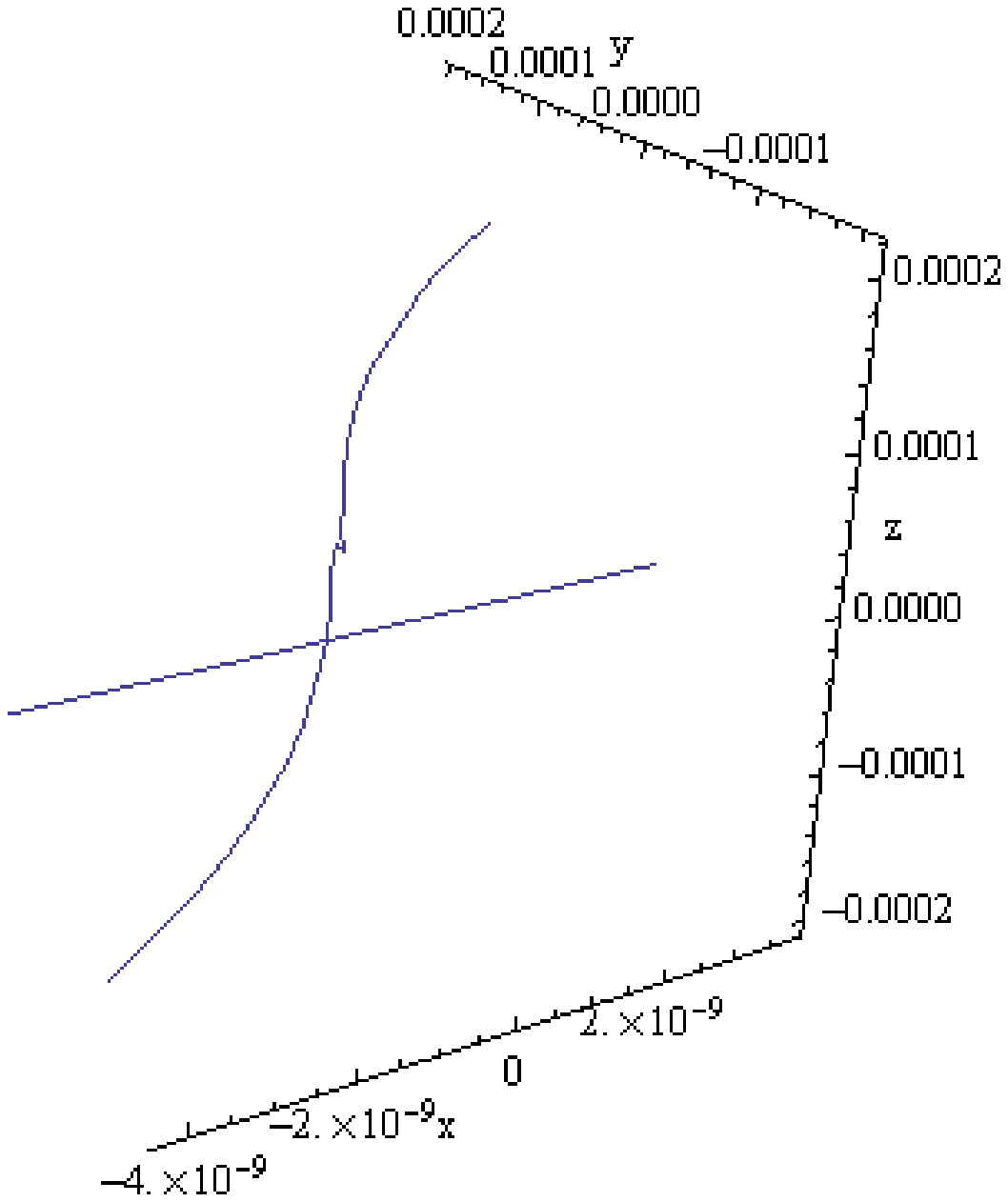}}
\caption{\small{Dynamics of the one-soliton solution $\textbf{m}$ of the modified coupled HF equation \eqref{eq4.22}. Panels (a)-(c)show the vector projections of $\textbf{m}$; (d) and (e) show dynamics of $\textbf{m}(x,1)$ and $\textbf{m}(-1.04,t)$ in space $R^{2+1}$. The parameters are chosen as $a=1/2,b=1.$}}
\label{fig1}
\end{figure*}
\begin{figure*}[t!]
\centering
\subfigure[] {\includegraphics[width=0.25\textwidth]{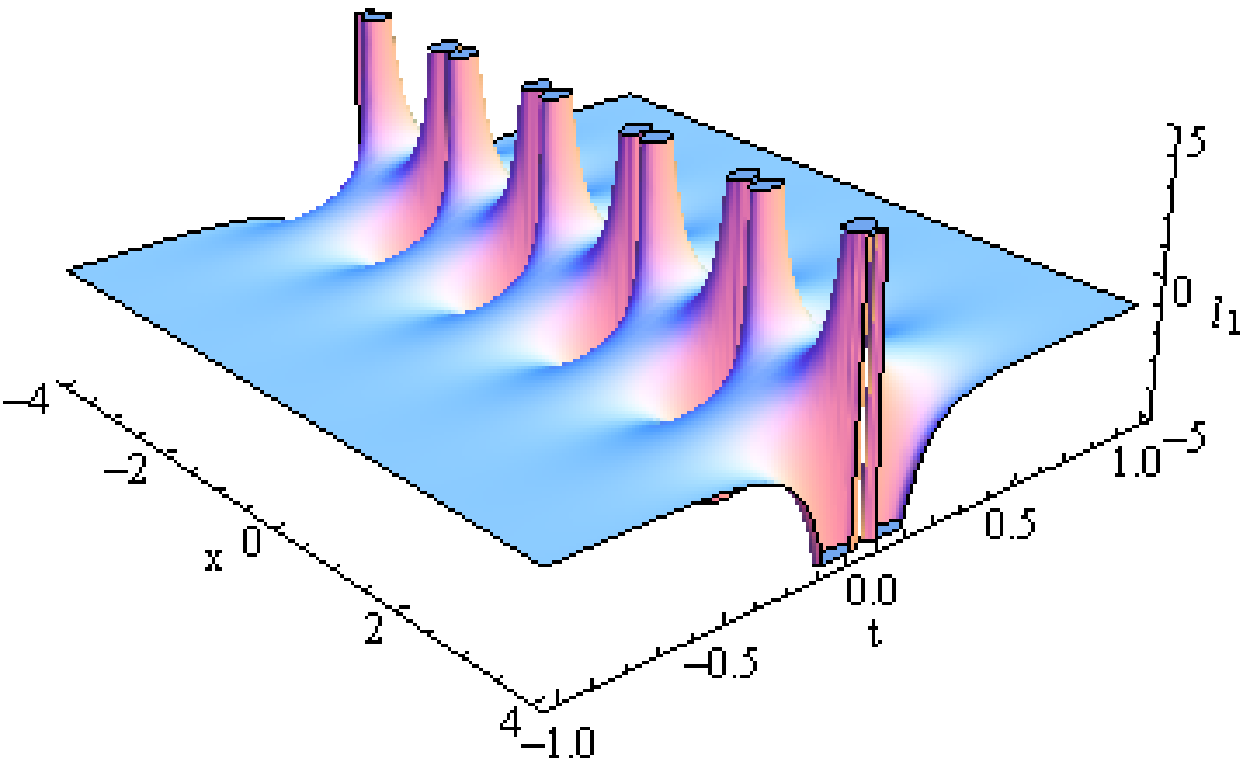}}\quad
\subfigure[] {\includegraphics[width=0.25\textwidth]{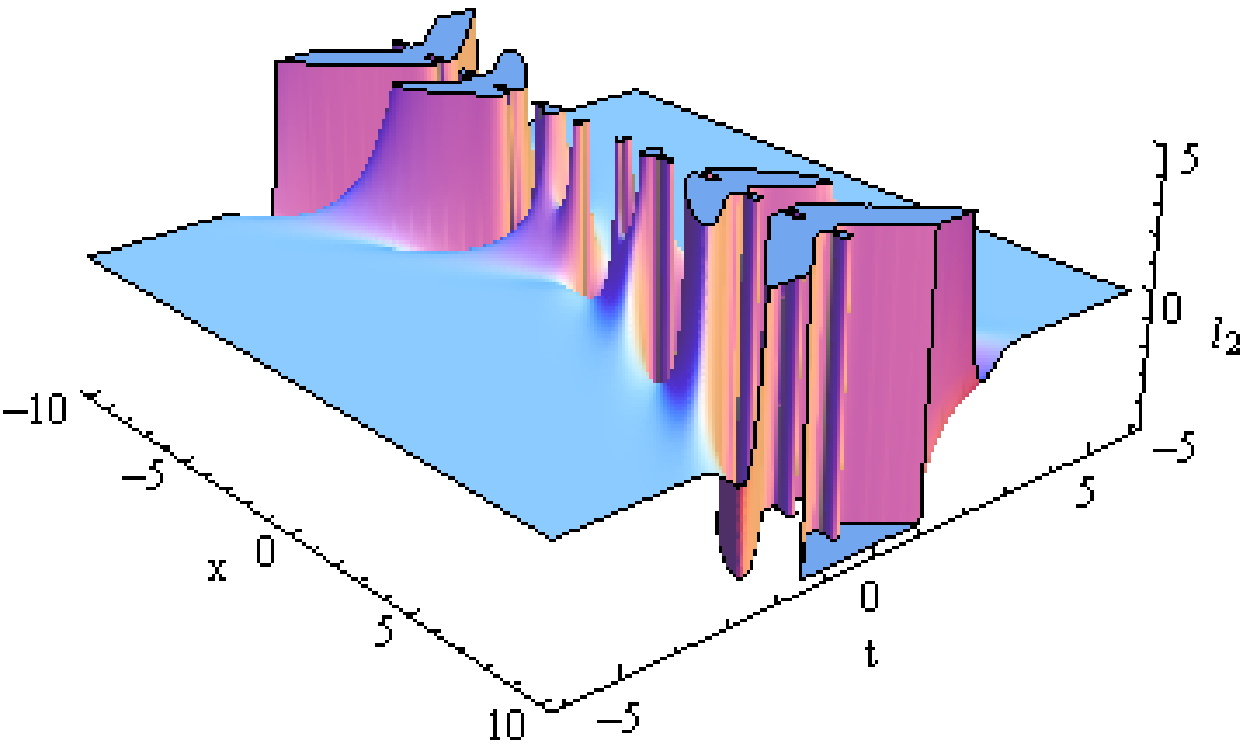}}\quad
\subfigure[] {\includegraphics[width=0.25\textwidth]{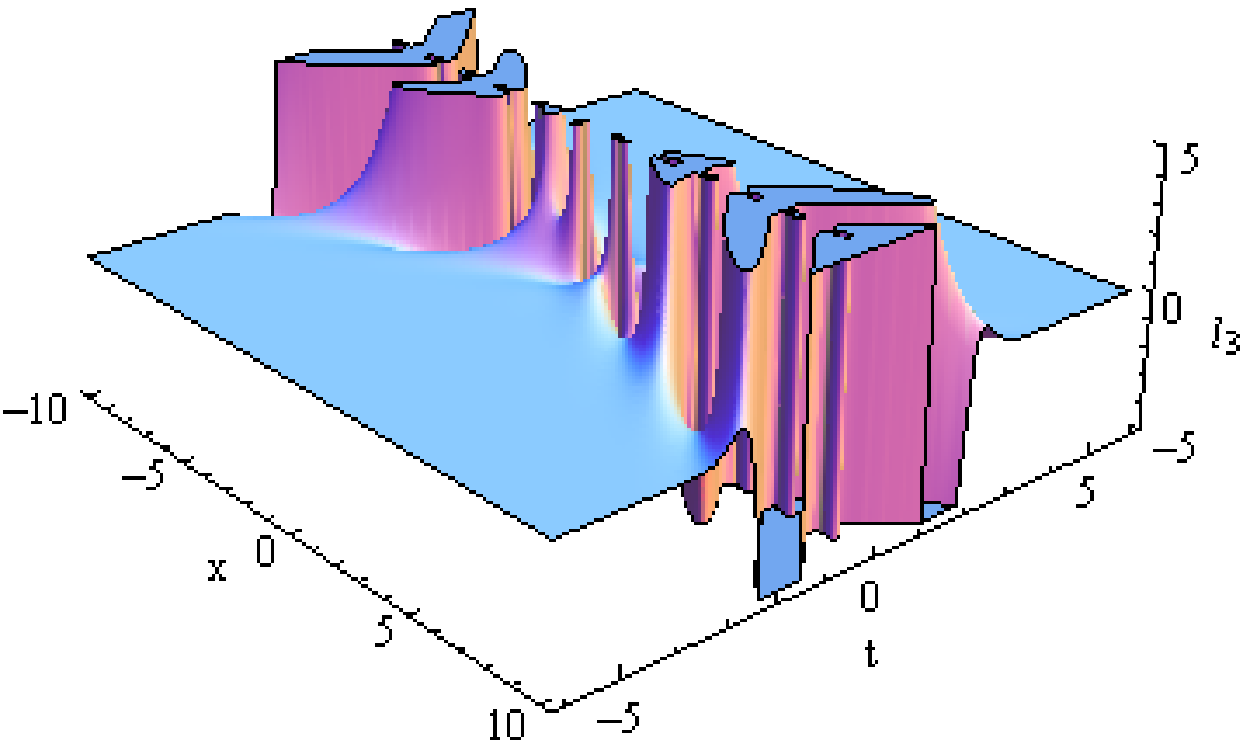}}\\
\subfigure[] {\includegraphics[width=0.23\textwidth]{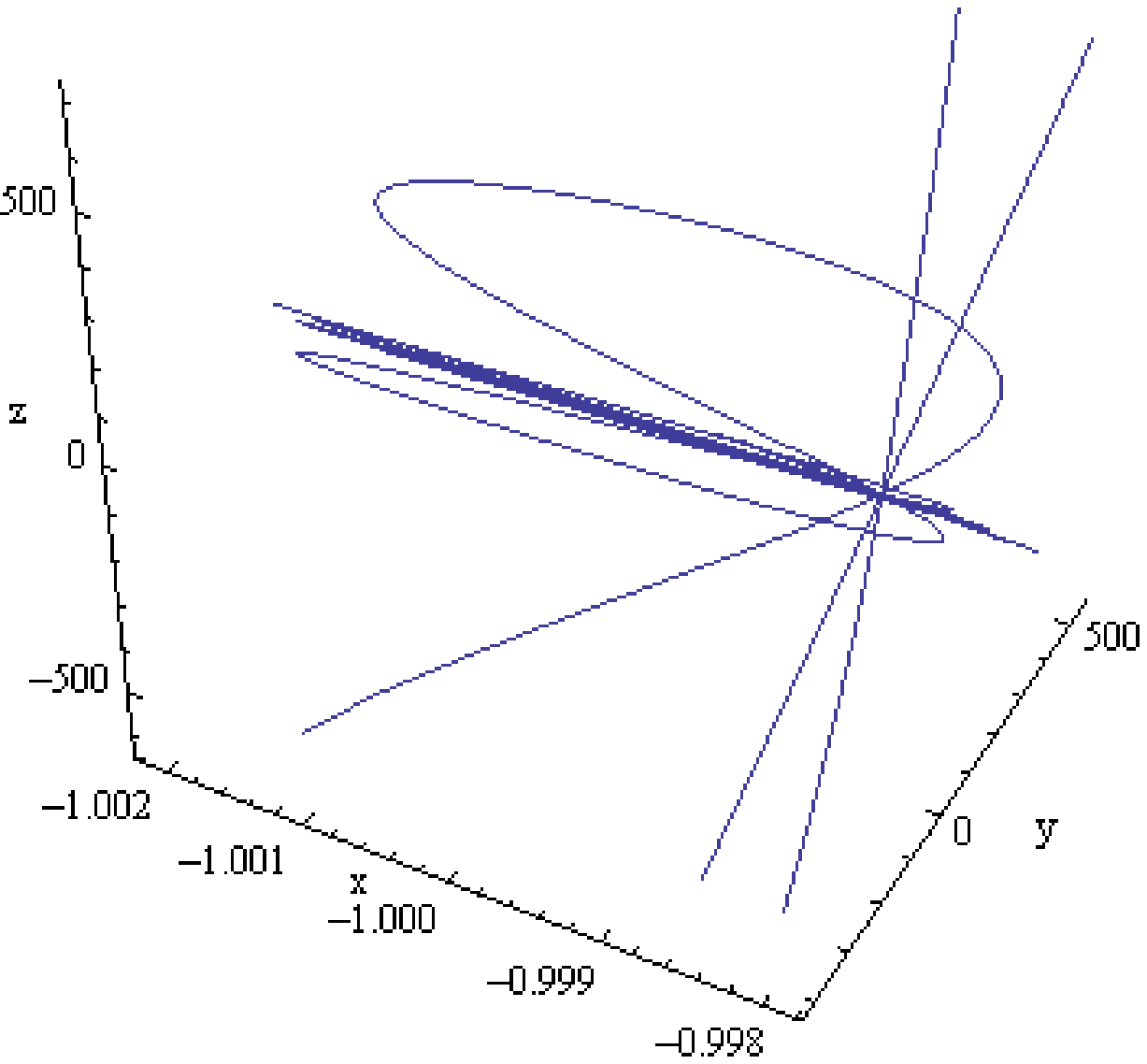}}\quad
\subfigure[] {\includegraphics[width=0.23\textwidth]{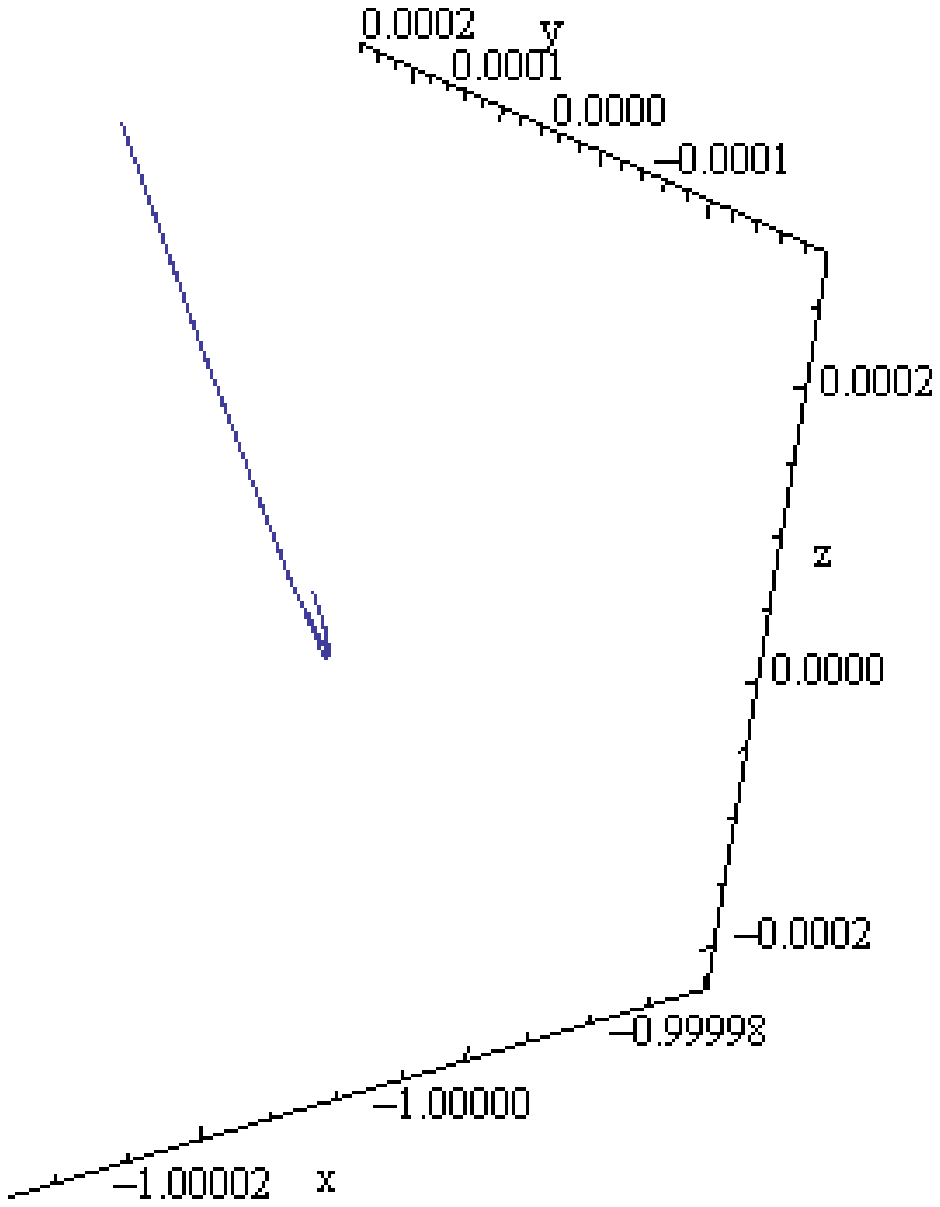}}
\caption{\small{Dynamics of $\textbf{l}$ in the one-soliton solution of the modified coupled HF equation \eqref{eq4.22}. Panels (a)-(c)show the vector projections of $\textbf{l}$; (d) and (e) show dynamics of $\textbf{l}(x,1)$ and $\textbf{l}(-1.04,t)$ in space $R^{2+1}$. The parameters are chosen as $a=1/2,b=1.$}}
\label{fig2}
\end{figure*}
\section{Conclusion}
\quad In this paper, we have established the relation between nonlocal NLS equation and coupled HF equation in the discrete and continuous cases. We have shown that the nonlocal discrete NLS$^+$ equation and the nonlocal discrete NLS$^-$ equation are, respectively, gauge equivalent to the discrete coupled HF equation and the modified discrete coupled HF equation. Under the continuous limit, the discrete coupled HF equation and the modified discrete coupled HF equation yield the corresponding coupled HF equation and the modified coupled HF equation. This means that the nonlocal NLS$^+$ equation and the nonlocal NLS$^-$ equation are gauge equivalent to the coupled HF equation and modified coupled HF equation. The exact solution of the modified coupled HF equations is obtained through the solution of nonlocal NLS$^-$ equation. Our results are significant for the deep understand of nonlocal NLS equation and its discrete version.

\vskip 15pt \noindent {\bf
Acknowledgements} \vskip 12pt

The work of ZNZ is supported by the National Natural Science
Foundation of China (NSFC) under grants 11271254, 11428102 and 11671255, and
in part by the Ministry of Economy and Competitiveness of Spain under
contracts MTM2012-37070 and MTM2016-80276-P, that of SFS by the NSFC under grant 11371323.

\small{

}
\end{document}